# Finding Communities in Site Web-Graphs and Citation Graphs

Antonis Sidiropoulos

asidirop@csd.auth.gr

Data Engineering Lab, Department of Informatics,
Aristotle University, Thessaloniki, 54124 Greece

**Abstract.** The Web is a typical example of a social network. One of the most intriguing features of the Web is its self-organization behavior, which is usually faced through the existence of communities. The discovery of the communities in a Web-graph can be used to improve the effectiveness of search engines, for purposes of prefetching, bibliographic citation ranking, spam detection, creation of road-maps and site graphs, etc. Correspondingly, a citation graph is also a social network which consists of communities. The identification of communities in citation graphs can enhance the bibliography search as well as the data-mining. In this paper we will present a fast algorithm which can identify the communities over a given unweighted/undirected graph. This graph may represent a web-graph or a citation graph.

## 1 Introduction

During the past decade the World Wide Web became the most popular network in the World. WWW grows with a very fast speed, thus the information that can be found through it is huge. Two are the main problems for the Web. How to find information and how to get it fast. For the former, several solutions have been presented over last years. From the ordering by the keyword frequency we have moved to the Link Analysis Ranking Algorithms (LAR). LAR Algorithms gave an admissible solution for the problem of searching. For the latter, the proxy servers and the Content Distribution Networks gave a breath to the problem of speed. However, the Web is still growing fast, the web-sites are huge and usually semi-automatically generated. So the above areas of research need to enhance their propositions.

On the other hand, the Web is a characteristic example of a social network [25, 19]. Socials networks are usually abstracted as graphs, comprised by vertices, edges (directed or not) and in some cases, with weights on these edges. Social networks have been studied long before the conception of the Web. Pioneering works for the characterization of the Web as a social network and for the study of its basic properties are due to the work of Barabasi and its colleagues [2]. Later, several studies investigated other aspects, like its scale-free nature [3, 1] , its growth [26, 4, 21, 18], etc.

One of the most intriguing features of the Web, and of other social networks as well, is its self-organization behavior, which is usually faced through the existence of *communities*. Groups of vertices that have high density of edges within them and lower density of edges between groups is a frequent informal definition of a community. The notion of a community is very useful from a practical perspective, because it can be used to improve the effectiveness of search engines [10], for purposes of prefetching [23], bibliographic citation ranking [22], spam detection [14], creation of road-maps and site graphs, etc.

In addition, the discovery of communities in citation graphs will also help to facilitate and enhance the bibliography search. For example, it could be possible to find relevant documents even if there are no common keywords and no direct citation between them. Also it will be possible to find authors with the same interests as well as communities of authors working on the same scientific domain.

## 2 Related Work

The notion of a Web community is not very strict; it is generally described as a substructure (subset of vertices) of a graph with dense linkage between the members of the community and sparse density outside the community. The existence of communities in the Web was first reported in [13]. The aforementioned qualitative definition though is not adequate when trying to devise algorithms for the determination of communities in Web graphs. Thus, we need sharper, quantitative definitions for the communities.

In order to provide such a quantitative definition, we need to introduce some "quantities" The basic quantity to consider is $d_i$, the degree of a generic node $i$ of the considered undirected graph $G$ (representing the examined network), which, in terms of its adjacency matrix $A[i, j]$, is $d_i = \sum_j A[i, j]$. If we consider a subgraph $V \subset G$, that node $i$ belongs to it, we can split the total degree $d$ in two quantities: $d_i(V) = d_i^{in}(V) + d_i^{out}(V)$. The first term of the summation denotes the number of edges that connect node $i$ to other nodes which belong to $V$, i.e., $d_i^{in} = \sum_{j \in V} A[i, j]$. The second term of the summation formula denotes the number of connections toward nodes in the rest of the graph, i.e., $d_i^{out} = \sum_{j \notin V} A[i, j]$. The first definition of communities is due to Flake [10, 9], who defined a community as the set of nodes $C(C \subset G)$ such that $d_i^{in}(C) > d_i^{out}(C) \forall i \in C$.

In general, we may give many different quantitative definitions of a community, which depend on the context of the application where it is developed. The structure of a community can be encountered at various scales in the Web. The most thoroughly investigated are the inter-site communities, which span several Web sites, and usually define broad thematic areas determined by a set of keywords, e.g., the 9/11 community [11]. The notions of compound documents [8, 7] and logical information units [24, 17] are closely related to the Web communities, but at a much smaller scale, being comprised by a handful of Web objects in a single site and thus they are intra-site communities.

We extend the notion of intra-site communities and proposed communities whose topic is much more generic than the topic of logical documents and their existence is determined by the density of the linkage among the pages that they are comprised of.

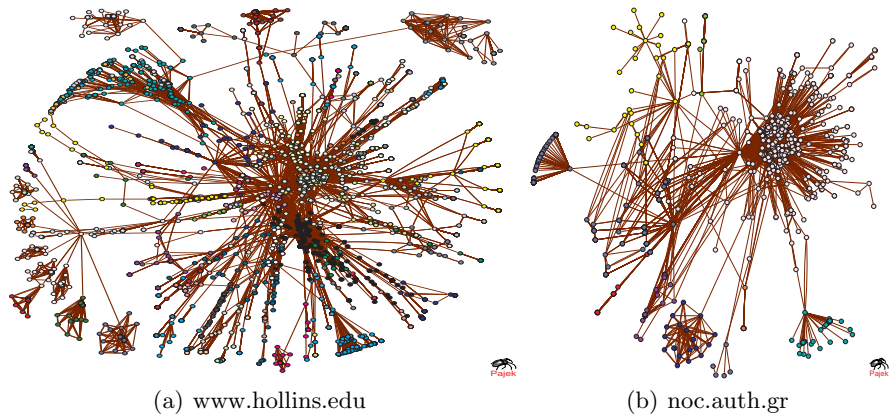

(a) www.hollins.edu        (b) noc.auth.gr

**Fig. 1.** Visualization of intra-site communities.

To support this claim, we examined several Web sites with a crawl available on the Web. As an intuitive step, we confirmed the existence of such communities using graph visualization[1]. As a sample, we present the drawing of the http://www.hollins.edu Web site (Figure 1(a), whose January 2004 webbot crawl was available on the Web. We can easily see the co-existence of compound documents (at the lower right corner), with compact node clusters (at the upper center), and less apparent clusters (at the upper right of the image). Also, the resulting image of the http://noc.auth.gr/ (as of Feb 2006) is illustrated at Figure 1(b).

In an analogous way, different type of communities do exist in an author citation (or collaboration) graph. An author may have worked in two institutions (working groups) so he should belong to two communities defined by the working groups. One working group may collaborate with another, so both working groups belong to a higher level group (hierarchical clustering). At the same time, only one person of a working group may collaborate with another one, so this person should belong to the cluster defined by his working group and in a higher level to the cluster defined by the second working group plus himself. So, in citation graphs different types of communities do exist at the same time.

In order to cover most of the above community cases, we give a weaker definition for communities than Flake et al. did. We define a community $C(C \subset G)$ such that [23]:

$$\frac{d^{out}(C)}{d^{in}(C)} < s \qquad (1)$$

where $d^{in}(C)$ is the number of links within the community and $d^{out}(C)$ is the number of links from members to non-members. We set the factor $s$ to 1, so we

---

have a basic 'community' but we can also set $s$ to any number less than 1 in order to find stronger communities.

The identification of communities is essentially a graph clustering procedure, that its goal is to identify mutually disjoint subsets of vertices, the communities. The discovery of optimal Web communities as well as any graph clustering is an NP-hard problem. Thus all the methods proposed rely on some properties of the graphs. Some methods evaluate only the local neighborhood of a vertex in order to decide whether it belongs to a specific community, whereas some other methods demand examination of the whole Web graph in order to discover such communities. No matter what method is selected to identify the communities, there always exists a trade-off associated with this task. This trade-off relates the 'quality' of the discovered communities, i.e., the density of linkage inside them, with the computational (time and/or resources) cost. In the rest of this section, we outline the most important methods for community discovery, namely bibliometric, spectral, maximum-flow and graph-theoretic techniques. The first family of methods exploits only local information, the second family is based on information from the whole graph, whereas the other two families can be tailored to use either local or global information or a combination of them.

**Bibliometric Methods** The bibliographic methods attempt to identify communities by searching for similarity between pairs of vertices. Thus, they have to answer the question 'Are these two pages similar'. To answer this question they need to define a similarity metric for the vertices. There are two two such metrics that are widely used. The first is the *co-citation coupling* and the second is the *bibliographic coupling*. Bibliographic techniques are relatively old and well-established techniques for the discovery of communities. More information on their application can be found in [6,12].

**Spectral Methods** The most popular spectral methods for community identification is HITS [16]. The HITS algorithm takes a subset of the Web graph based on a keyword match. Then it extents this set by adding nodes that are 2 links away from any node already in the subset. If $A$ is the adjacency matrix of the subgraph, the matrix products $A^T A$ and $AA^T$ are symmetric and definitely positive. Each of them will have the property of being identical the left and the right eigenvectors (because of symmetry) and that the first eigenvector will have all positive components (with positive eigenvalue). These subsequent eigenvectors can be used to distinguish pages into different communities in a manner related to the spectral graph partitioning. using a method such that, it was found that the non-maximal eigenvectors can be used to split pages from a base set into multiple communities that contain similar text but they are dissimilar in meaning.

**Maximum-Flow Methods** Flake et al. in a series of papers used the concept of max-flow/min-cut in order to discover communities in Web graphs. The algorithm proposed [10,11] works as following. Its input is a graph G, a set of 'seed'

Web pages S, and a single (user-defined) parameter $\alpha$. The procedure creates a new graph, $G_\alpha$, that has one artificial vertex $t$. The sink vertex, t, is connected to all original vertices with a small capacity specified by $\alpha$. After constructing $G_\alpha$, the procedure calls min-cut for randomly selected source vertex $s$ to $t$ and uses the resulting residual graph to return the portion of R that remains connected to s. This connected component is guaranteed to be a community.

**Graph-Theoretic Methods** We saw earlier that the bibliometric methods try to identify the "strongest" edges in order to insert the adjacent vertices into a community. Girvan and Newman ([15]) took the opposite approach, following a graph-theoretic way. Instead of trying to construct a measure that tells us which edges are the most central to the communities, instead they focused on those edges that are least central, the edges that are most "between" the communities. Rather than constructing communities by adding the strongest edges to an initially empty vertex set, they construct them by progressively removing edges from the original graph. They exploited the vertex betweenness, which had been studied in the past as a measure of the centrality and influence of nodes in networks. The betweenness centrality of a vertex i is defined as the number of the shortest paths between pairs of other vertices that run through i. They proposed a simple algorithm and its steps are the following: 1. Calculate the betweenness for all edges in the network. 2. Remove the edge with the highest betweenness. 3. Recalculate betweenness for all edges that are affected by the removal. 4. Repeat from step 2 until no edges remain. With this process the graph is gradually being disconnected revealing any communities.

## 3    Motivation & Contribution

The above described techniques have some strong as well as some weak points. Starting with the *bibliographic methods* we can say that they can only be applied to a citation graph and not to a web-graph because they usually need specific information about vertex relationship, i.e. co-authors.

The *Spectral methods* can only be used when keyword information is available over the graph. Also, they are not capable of finding the communities of the graph but only a community related to a keyword search. This means that it cannot be applied when "keyword" info is not available or we do not a-priori know the "keyword" related communities.

The *maximum-flow methods* have some major disadvantages. The first disadvantage is that they are based on a very "strict" definition for communities. If our graph has not such strict communities the method is unable to find any. The second one is that they are mainly used in order to find inter-site communities by removing the intra-site links. The existence of intra-site links practically invalidates the results. On the other hand the commutation time is large, and it is based on several decisions that must be made during the implementation of the algorithm. So a lot of variations exist. As the variations get better performance, the computation time increases dramatically. The most important disadvantage of these methods, is the existence of the factor $\alpha$ which must be set manually and there is no rule for seting it relatively to the graph characteristics. The only method is to perform a binary search for a good value of $\alpha$ which practically means several failed tries in order to get a clustering.

Finally the *Graph-theoretic methods*, as we will describe later, require high memory usage as well as long computation time.

In addition to the above, in the real world, a web-page may not belong strictly to one community, but to more than one. Likewise, a web page may not belong to any community. Thus the set of communities $C_1, C_2, ..., C_k$ may $\bigcup_i C_i \subset G$ and not always $\bigcup_i C_i = G$. There also may exist intersections between the communities. So it may exist $i, j (i \neq j)$ such that $C_i \cap C_j \neq \emptyset$. This is our main theoretical difference with all the rest methods.

So, our method searches for a set of clusters $C = \{C_1, C_2, ...\}$ such that $\forall C_i \in C : \frac{d^{out}(C_i)}{d^{in}(C_i)} < s$. $s$ is user defined but its default value is 1. There may exist infinite clusterings with this property. We focus to a clustering that minimizes the expression:

$$Q_C = \frac{1}{|C|} \sum_{\forall C_i \in C} \frac{d^{out}(C_i)}{d^{in}(C_i)} \tag{2}$$

## 4 Proposed Method

The Betweenness Centrality is a way of showing how central is every node of a graph $G$. Many algorithms have been presented in the bibliography for the calculation of the Betweenness Centrality. The smartest of them is [5] with computational complexity $\mathcal{O}(nm)$ and memory requirement of $\mathcal{O}(m + n)$, whre $n$ is the number of nodes and $m$ is the number of edges.

### 4.1 Clustering Using Betweenness Centrality

The notion of CB is used in paper [20] for the clustering of a graph . In this paper, the CB is calculated for each edge of the graph. At each step, the edge $e$ with the highest CB is removed from the graph and the CB is recalculated for some of the edges. In other words, all the shortest paths that contained the erased edges $e$ are recalculated and the CB is recomputed for all edges. This procedure is repeated until we get groups that are not connected to each other (connected components). The complexity of this algorithm is high in time $\mathcal{O}(n^3)$, since the CB is recalculated in every step of the algorithm. Moreover, it has expensive memory requirement $\mathcal{O}(n^2)$, since we have to store all the shortest paths for all the node pairs. This forbids the use of this algorithm for large graphs and specially for dense ones, because the memory requirement becomes huge. The authors of [20] mention that using our age computers (2003) the described method can be applied in graphs of about 10000 nodes.

The conclusion of the above paper is that vertices with high CB are near to the borders of the clusters as well as edges with high CB are inter-cluster edges.

On the other hand vertices and edges with low CB reside at the center of the clusters or simply are not connected to other clusters.

The above claim is not true when a part of our graph has a tree-like structure. A tree-like part of the graph means that in this sub-graph there are no cycles and the number of edges is equal to the number of vertices. These parts look like graph tails. The existence of such parts in our graph inflect the previous statement. All the nodes in such a sub-graph have high CB but they clearly consist a cluster. We assume that in a Web-graph these parts do not cover a big ratio of the graph. We refer to these tree-like parts of the graph as *graph tails*.

Practically, tree-like tails in a web graph represent virtual documents which do not have links pointing out of the document. So, they may consist of independent clusters of our graph or they may be members of other clusters.

### 4.2 Our Method

The algorithm CBC (Clustering with Betweenness Centrality) begins with the knowledge that the nodes with the lower CB are members of clusters and they are not connected directly to other clusters. Initially we remove the *graph tails* from graph $G$ as described before, resulting to a new graph $G'$. We compute the centrality based on the $G'$. The rest of the procedure is depicted in Figure 2, where $C$ is initially an empty set of clusters.

**Clique Formation** The nodes with the lower CB are the cluster *kernels*. So, we can build some initial small clusters around them. This is depicted as pseudo code in Figure 3. These small clusters are the graph *Cliques*. Note here that our term *Cliques* if different than it is used in the bibliography where usaly means a fully connected sub-graph. The *Cliques* for us are the areas around the it kernel nodes. The clique size may vary and this is a function of how dense or sparse the graph is. For a dense graph the cliques consist of all the nodes that are directly connected to the kernel node. For sparse graphs the cliques may be larger. The optimal clique size, can not be known apriori, since the graph may be dense in some areas, but sparse in other ones. The first time that *InitiateCliques* will called, the *Maximum Clique Size* parameter is set to zero. Thus, the Cliques that will be built during the first iteration of the algorithm have a diameter of two. Note here that if an Initial Clique has a size of only one or two nodes, it is

```
function InitClustering(graph G,G', clustering C, int max_clique_size) {
    InitiateCliques(G',C,max_clique_size);
    ExtentTailedClusters(G,C); // Add the tails of size 1
                               // to the cliques that they are connected
    MakeTailedClusters(G,C);   // Make the tree-like tails independent Cliques
    MergeTailedClusters(G,C);  // Merge tail-clusters until reach the max-cluster-size
}
```

**Fig. 2.** Init Clustering

```
function InitiateCliques(graph G,clustering C) {          function ExtentClique(graph G,
   static max_clique_size=sqrt(G.n_nodes);                  cluster c, clustering C,
   for all nodes n in G ordered by CB desc {                int max_size) {
      cluster c;                                             if(c.n_nodes>max_size) {
      if(n belongs to any c in C) {                             return false
         next;                                               }
      }                                                      Extent c using BFS until the
      c = {n};                                               size of 2*max_size
      for all p neighbors of n {                             return true
            if(not p belongs to any c in C) {             }
               c.add(p);
            }
      }
      C.add(c)
      while(ExtentClique(G,c,C,max_clique_size)) {;}
   }
   max_clique_size=next value;
}
```

**Fig. 3.** Initiate Cliques

being erased and ignored. This means that after the first step we may have some *orphan* nodes. These will be nodes with high CB and usually located between the clusters. The computation cost of this step is a linear function of the size of the graph: $\mathcal{O}(n)$.

**Clique Merging** Having a set of cliques the next step is to merge them in order to construct correlated clusters. The cliques may not be correlated. The pseudo-code is presented in Figure 4. Assuming that in the previous step we found $l$ number of clusters (cliques), in function *Merge* we build a $l \times l$ matrix B. Each element $B[i,j]$ corresponds to the number of edges from cluster $C_i$ to cluster $C_j$. The diagonal elements $B[i,i]$ correspond to the number of internal edges of the clique. The matrix $B$ is obviously symmetric. Note here that if a node $x$ belongs to clusters $C_i$ and $C_j$, and there is an edge $x \rightarrow y(y \in C_i)$, then this edge counts once for $B[i,i]$ since it is an internal edge, but it also counts for $B[i,j]$, since $y$ belongs also to $C_j$. So, the sum of a row of matrix $B$ is not equal to the total number of edges.

This merging step of the algorithm consists of several iterations. In each iteration one merge is performed. The iterations stop when there is not any other mergable pair. The pair that will be merged is the pair with the maximum $B[i,j]/B[i,i]$. The conditions for a merge may vary and they depend on the user parameters if there are any. A parameter may be the factor $s$, which is mentioned in the definition of a community. So, in this step we check every pair of the clusters (cliques) and we select the best one for merging. A merge cannot be done if the two clusters are already correlated or their union is greater than the *maximum cluster size* that the user may have set. A merge of clusters $C_i, C_j$ can be done if $B[i,j] > \sum_k B[i,k]/2$ or if at least one of the $C_i, C_j$ is not correlated or if $B[i,j]/B[i,i] >= s$ and we cannot find any better pair.

If the initial number of cliques is $l$, then the maximum number of iterations that will be executed is $l$. Each iteration checks every pair of cliques, so the time

```
function ClusterMerge(graph G,clustering C...) {
    while(! ok) {
        remove_from_cliques_to_make_stronger
        Merge(G,C);
        if(c->best_quality==0) {manage_subsets(c);}
        delete_the_worst(C);  // Delete a cluster if does not fulfil parameters
                              // Or it is not correlated.
        add_orphans_to_cliques(G,C);
        ok=check(C);
        if(!ok){
            InitClustering(G,C);
        }
    }
}
```

**Fig. 4.** Cluster Merge

that is needed is $l^2 + (l-1)^2 + ...$ which means that the complexity is $\mathcal{O}(l^3)$. The value of $l$ depends on the graph characteristics, but it is a function of $\sqrt{n}$, where $n$ is the number of nodes in our graph[2]. So the time complexity, with respect to the number of nodes is $\mathcal{O}(n\sqrt{n})$ in the worst case. The memory space requirements is a matrix $l \times l$ which means $\mathcal{O}(l^2)$, that is near to $\mathcal{O}(n)$.

In Figure 4 there are various steps in order to improve quality and/or speed depending on our needs. For example, the procedure named *RemoveFromCliques ToMakeStronger* can be used for producing *Flake* compatible clusters. Its default behavior is to check for all the nodes (in $\mathcal{O}(n)$ time) the number of the links that they have to each cluster. So, a node may change cluster if it does not fulfil our constraints. The function *ManageSubsets* is used to remove any clusters that are subsets of other clusters. It will be called only for speed optimization. Finally we may add orphan nodes to clusters, even if the resulting clustering is not better than the current one. This will lead the algorithm to minimize the number of orphan nodes but the resulting quality factor of Equation 2 $Q_C$ will be worst.

Finally the clustering is checked if it fulfils our constraints. In not, we re-initialize into cliques the nodes that remained as orphans. Each time that the *InitClustering* is called, we use a new value for the factor *max clique size*. In our implementation the sequence of values is: $\sqrt{n}, \sqrt{n}/2, 2\sqrt{n}, \sqrt{n}/3, 3\sqrt{n}, ....$ In most cases that we faced during experimentation, the clustering is computed in two repetitions of function *InitClustering* and in a few cases in four. Of course if the cluster sizes vary, we expect that more repetitions will be needed.

## 5   Method Evaluation

As mentioned in the previous section, an analogous idea is the one described in [20]. The major disadvantage of that algorithm is the huge complexity, since every step of the algorithm it requires recomputation of the centrality. Although the computation is incremental, the time and space complexity is very high. In this work we will not compare against this algorithm, since the difference in the complexity and the memory requirements is obvious.

---

[2] Given that we have initially set the parameter *max clique size* to $\sqrt{n}$.

## 5.1 Evaluation Dataset

The evaluation Dataset consists of real and synthetic Web-graphs. The real Dataset includes the web sites of: noc.auth.gr (as of Feb 2006), www.hollings.edu (as of Jan 2004) and www.unisef.org (as of Jan 2006).

The synthetic web-graphs were generated with the FWgen tool. The parameters that must be given to FWgen are five: (a) number of nodes, (b) number of edges or density related to the respective fully connected graph, (c) number of clusters to generate, (d) skew level, which reflects the relative sizes of the generated clusters, and finally (e) the assortativity factor, which gives the percentage of the edges that will be intra-community edges. The values that are meaningful for assortativity are greater than 50%. The higher the assortativity is, the stronger clusters are produced. If the assortativity is 100% then the generated clusters will be disconnected to each other.

The generator creates two files. The first one is the graph and the second one records the produced clusters so that we can compare the clustering of the CBC to the "generated optimal". Since the generation of the edges follows random decisions, the "generated optimal" clustering may not be identical to the "absolutely optimal". With the term "absolutely optimal" we mean the clustering that minimizes the factor $Q_C$. This will be shown later in this section.

## 5.2 Evaluation Method

The Method is evaluated for two dimensions: quality and speed. The evaluation for quality could be done only with the synthetic graphs, for which we know apriori the clusters that are constructed. So, we count the distance of our method from the optimal clustering by using a distance metric explained in Appendix A. We also count the value $Q_C$ of Equation 2. The smaller this value is, the better clustering is produced. On the other hand the evaluation of the clustering speed is trivial. We count the real-time of the algorithm execution (CPU time occupied by the process). For the real dataset we present statistical results of the clustering.

## 5.3 Experiments

In Figures 5(a,b) we present the speed performance of our algorithm. The presented "CPU Time" is reported by the unix kernel by using the system call *times()* and represents the time that the process remained in CPU. The green line (using diagonal crosses for points) represents the time needed to compute the centrality. The red line (with cross points) represents the time needed by our algorithm to compute the clustering. Finally the blue line (with star points) represents the summation of all the previous. As we can see, our algorithm needs much less time than the centrality betweenness, which is proved to have $\mathcal{O}(mn)$

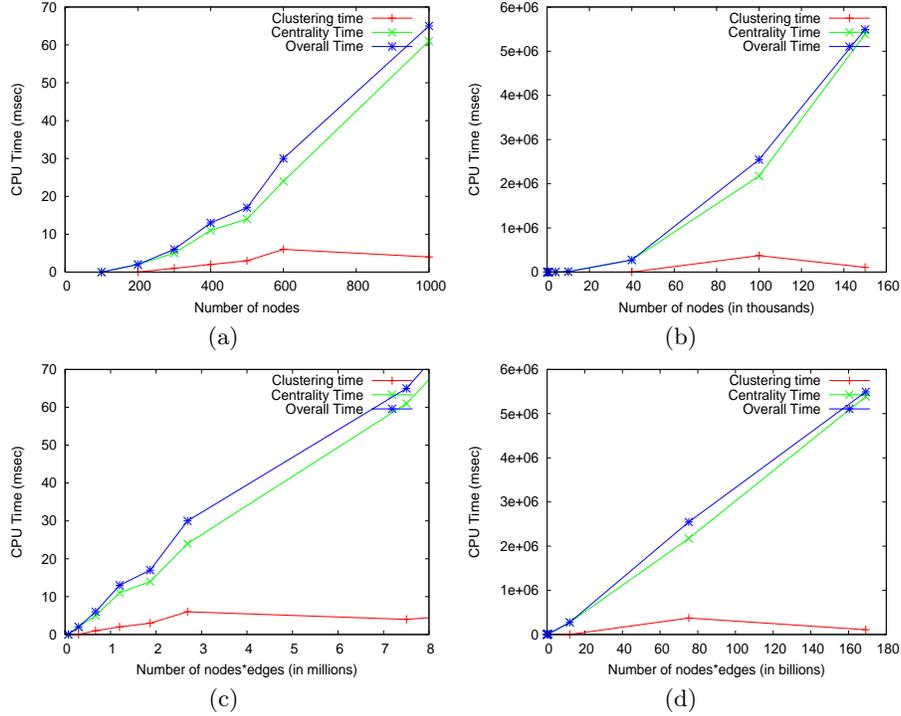

**Fig. 5.** Graphs attributes: Nodes: $n$, Edges: $15 * n$, clusters: 5 ($n < 1000$), 10 ($1000 \leq n < 10000$), 100 ($10000 \leq n$), assortativity: 0.85, skew:0.1

complexity. In Figures 5(c,d), we verify that the CB computation is linear to $n * m$, so our time measurements are correct. Thus, if we use a centrality approximation algorithm, we will be able to cluster really huge graphs consisting of a lot more than 200000 nodes.

In Figure 6 we present the results of the clustering that are related to the assortativity parameter. Figure 6(a) shows the distance of our clustering from the "generated optimal". The distance is computed by using the method presented in Appendix A. The red line (with the cross points) stands for our BCB algorithm, while the green line (with the diagonal cross points) stands for our algorithm that uses the option *minimize orphan nodes* set. As we can see, the *minimize orphan nodes* version gives a clustering closer to the "generated optimal". This happens because in the "generated optimal" clustering there are no orphan nodes. The distance from the "optimal" is in the worst case 1% and it converges to zero as the clusters become stronger. On the other hand, in Figure 6(b) we present the quality of the clustering. It is expressed with the factor $Q_C$ that is defined in Equation 2. It is obvious that when the clusters are strong the quality of the clustering is better. Hereby we must note that both our CBC versions keep the quality very close to the "generated optimal" clustering and they are always better than it. This is due to the fact that the generator does not produces optimal clusters, but if they do exist in the graph, our algorithm is able to find them. This explains the fact that in Figure 6(a) the distance from the "generated optimal" clustering is not zero.

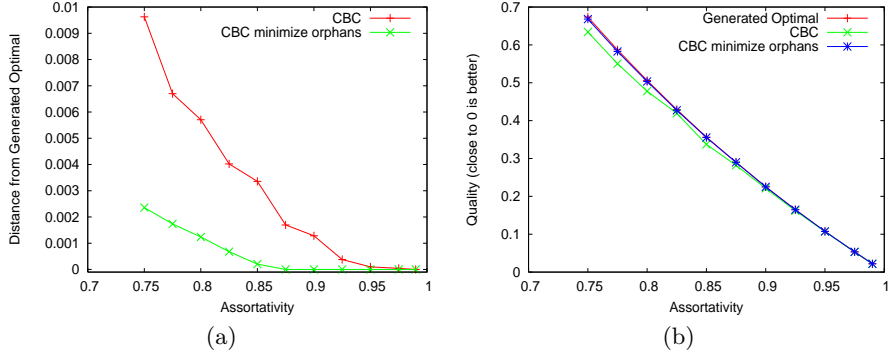

**Fig. 6.** Graphs attributes: Nodes: 4000, Edges: 30000, clusters: 10, skew:0.10

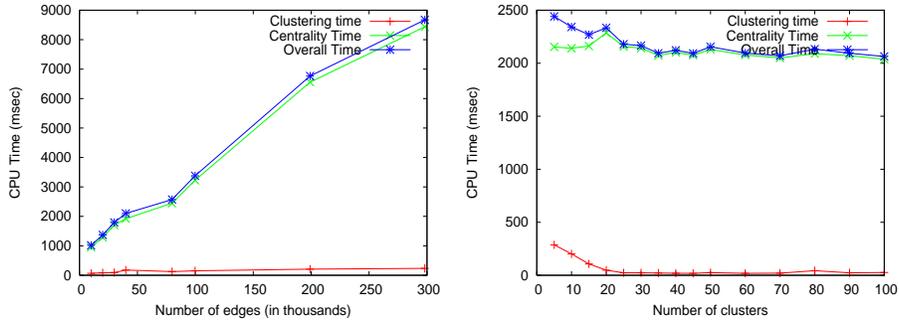

**Fig. 7.** Attributes: Nodes: 4000, clusters: 10, assortativity: 0.90, skew:0.10

**Fig. 8.** Attributes: Nodes: 5000, edges: 37500-39000, assortativity: 0.90, skew:0.10

In Figure 7, we keep the graph characteristics stable and we change the number of edges. As we can see the time that is needed for the clustering remains small. Finally, in Figure 8, in x-axes we change the number of clusters. It is shown that when the clusters are few, the required time is higher from the one that is needed than more clusters. This is due to the fact that more merge operations must be performed.

In Table 1, we present summarization of the results for the real web-graphs. The first 3 columns define the graph. Columns MC and MS stand for the user parameters *maximum cluster size* and *minimum cluster size*. Next column (Clusters) contains the number of clusters that have been found during each run. $Q_C$ is the resulting Quality (Equation 2) of the clustering. Column "Or" denotes the number of the orphan nodes that remained in the graph. Finally, the *Drate* column shows the percentage of the nodes that belong to more than one cluster. It is computed as:

$$Drate = \frac{\sum_{\forall i \in G} N(i)}{N_C}$$

| Site | Nodes | Edges | MC | MS | Clusters | $Q_C$ | *Drate* | Or |
|---|---|---|---|---|---|---|---|---|
| noc.auth | 955 | 5620 | 50% | 5 | 3 | 0.994 | 1.00 | 486 |
| noc.auth | 955 | 5620 | 80% | 5 | 10 | 0.332 | 1.04 | 3 |
| hollins | 4487 | 16373 | 50% | 10 | 64 | 0.204 | 1.05 | 170 |
| unicef | 56852 | 749666 | 30% | 10 | 142 | 0.544 | 1.14 | 67 |
| unicef | 56852 | 749666 | 30% | 1000 | 12 | 0.206 | 1.12 | 91 |

**Table 1.** Results over real web-graphs.

| Nodes | $C^{in}$ | $C^{out}$ | $C^{out}/C^{in}$ |
|---|---|---|---|
| 496 | 4846 | 143 | 0.0295089 |
| 272 | 322 | 154 | 0.478261 |
| 62 | 61 | 2 | 0.0327869 |
| 46 | 83 | 46 | 0.554217 |
| 44 | 47 | 33 | 0.702128 |
| 19 | 59 | 1 | 0.0169492 |
| 16 | 15 | 2 | 0.133333 |
| 16 | 15 | 2 | 0.133333 |
| 13 | 11 | 6 | 0.545455 |
| 12 | 10 | 7 | 0.7 |

**Table 2.** Cl. Results of http://noc.auth.gr.

where $N(i)$ is the number of clusters that node $i$ belongs to, and $N_C$ is the number of nodes that belong to at least one cluster. A value of 1 for *Drate* means that all nodes belong to exactly one cluster.

Since the possible clusters that may have $d^{out}/d^{in} < s$ may be infinite, we must somehow focus in some of them. For this reason, our implementation takes two parameters. The first one is the *minimum cluster size* (MS in Table 1) and the second one is the *maximum cluster size* relatively to the graph size (MC in Table 1). It is obvious that these two params affect the results. For example the first try to cluster the site noc.auth.gr used the default value 50% as MC. This caused the algorithm to leave a lot of orphaned nodes. The second run was by using MC=80%. This produced a big cluster of 496 nodes (Table 2), which is grater than the half of the graph and as we saw it could not be splited into smaller clusters. The results of this clustering are also visualized in Figure 1(b) where each cluster is presented with different color. Unfortunately, it is impossible to visualize the nodes which belong to more than one cluster since one node can have only one color. These nodes get the color from a randomly selected cluster among the ones they belong. Full results for all these experiments are available at *http://delab.csd.auth.gr/˜asidirop/clustering*.

## 6 Conclusion & Future Work

In this paper we made an overview of clustering methods. We also presented our method called CBC, as well as experiments over both synthetic and real datasets. The experiments show, as we expected, that this method is very fast and can be used in order to cluster huge web-graphs. As it is shown, the slow part of the method is the computation of the betweenness centrality. As a future work we plan to use a centrality approximation algorithm to test the clustering speed and the quality performance. This method is also being tested for prefetching methods over a content distribution network[23].

## Appendix A.    Clustering Comparison

In this section we will give a definition for a distance function for 2 clusterings.

We define the function $\mathcal{N}(n, c)$ to be 1, if node $n$ belongs to cluster $c$ and 0 otherwise. Also, function $\mathcal{K}(n, \mathcal{C})$ gives the set of clusters that node $n$ belongs to. The number of clusters that a node belongs to may be zero, one or any other number in the range of $[0..|\mathcal{C}|]$ when the node belongs to more than one clusters.

The similarity $S$ of a node $n_1$ to node $n_2$, given a set of clusters $\mathcal{C}$, is set to be the percent of the occurrences of node $n_2$ in $\mathcal{K}(n_1, \mathcal{C})$.

$$S(n_1, n_2, \mathcal{C}) = \frac{\sum\limits_{\forall c \in \mathcal{K}(n_1, \mathcal{C})} \mathcal{N}(n_2, c)}{|\mathcal{K}(n_1, \mathcal{C})|}$$

In the case where a node can belong only to one cluster, then the function $S$ will get the value of 1 if the two nodes are members of the same cluster, and 0 otherwise. In the general case that a node can belong to more than one cluster, then when two nodes always resize in the same clusters, $S$ will be 1. If two nodes never resize in the same clusters, $S$ will be 0. $S$ will be 0.5, if the first node belongs to 2 clusters and the second node belongs to only one of them, etc. Note here that it could be $S(n_1, n_2, \mathcal{C}) \neq S(n_2, n_1, \mathcal{C})$, in the case that the nodes are able to belong to different number of clusters. Iff a node can belong to only one cluster, then $S(n_1, n_2, \mathcal{C}) = S(n_2, n_1, \mathcal{C})$

$$\mathcal{D}(\mathcal{C}_A, \mathcal{C}_B, \mathcal{G}) = \frac{\sum\limits_{\forall n_1 \in \mathcal{V}} \sum\limits_{\forall n_2 \in \mathcal{V}} |S(n_1, n_2, \mathcal{C}_A) - S(n_1, n_2, \mathcal{C}_B)|}{|V|(|V| - 1|)} \quad (3)$$

In order to be able to compare 2 clusterings, we will give the definition of the equation $\mathcal{D}(\mathcal{C}_A, \mathcal{C}_B, \mathcal{G})$ (3), where $\mathcal{C}_A$ is the set of clusters created by method A and $\mathcal{C}_B$ is the set of clusters created by method B over the graph $\mathcal{G} = (V, E)$. $\mathcal{D}$ is calculated as the average value of the similarity differences in these two clusterings for every nodes pair. $\mathcal{D}$ is normalized in the scale of $[0..1]$. Iff the two clusterings $\mathcal{C}_A$ and $\mathcal{C}_B$ are equal, then $\mathcal{D}$ will be 0. The worst and only case that $\mathcal{D}$ may be 1, is the following: $\mathcal{C}_A$ has only one cluster and all the nodes of graph $\mathcal{G}$ belong to that cluster. $\mathcal{C}_B$ has as many clusters as the number of nodes and every cluster consists of only one node.

**Acknowledgments:** We deeply acknowledge Ulrik Brandes [5] help by providing us the implementation of the Betweenness Centrality Computation. Additionally, we are really grateful to Dimitrios Katsaros [23] for his inspired ideas as well as for the implementation of the graph generator FWgen.